\documentclass[a4paper]{article}

\usepackage{atmohead2013}
\usepackage[english]{babel}
\usepackage{xspace}

\def \pao      {Pierre Auger Observatory\xspace}
\def \mal      {Malarg\"{u}e\xspace}
\def \rnd      {R\&D\xspace}
\def \gcm      {g\,cm$^{-2}$\xspace}
\def \degree   {$^{\circ}$\xspace}
\def \xmax     {$X_{\rm max}$\xspace}
\def \us       {\char`\_}

\title{Global Atmospheric Models for Cosmic Ray Detectors}

\shorttitle{Global Atmospheric Models for Cosmic Ray Detectors}

\authors{
Martin Will$^{1,*}$
for the Pierre Auger Collaboration$^{2}$
}

\afiliations{
$^1$ Karlsruher Institut f\"{u}r Technologie, Institut f\"{u}r Kernphysik, Karlsruhe, Germany \\
$^2$ Full author list: http://www.auger.org/archive/authors\texttt{\us}2013\texttt{\us}06.html \\
\scriptsize{
$^{*}$ now at: Institut de Fisica d'Altes Energies, Bellaterra, Barcelona, Spain \\
}
}

\email{martin.will@kit.edu}

\abstract{
The knowledge of atmospheric parameters --~such as temperature, pressure, and
humidity~-- is very important for a proper reconstruction of air showers,
especially with the fluorescence technique. The Global Data Assimilation System
(GDAS) provides altitude-dependent profiles of these state variables of the
atmosphere and several more. Every three hours, a new data set on 23~constant
pressure level plus an additional surface values is available for the entire
globe. These GDAS data are now used in the standard air shower reconstruction of
the \pao. The validity of the data was verified by comparisons with monthly
models that were averaged from on-site meteorological radio soundings and
weather station measurements obtained at the Observatory in \mal. Comparisons of
reconstructions using the GDAS data and the monthly models are also presented.
Since GDAS is a global model, the data can potentially be used for other cosmic
and gamma ray detectors. Several studies were already performed or are underway
for several locations worldwide. As an example, a study performed in Colorado as
part of an Atmospheric \rnd for a possible future cosmic ray observatory is
presented.
}


\begin{document}
\maketitle

\section{Introduction\label{sec:introduction}}

A cosmic ray particle entering the atmosphere can initiate an extensive air
shower. The secondary shower particles excite nitrogen molecules in the air
which emit a characteristic, isotropic emission in the UV range as part of their
de-excitation process. The light can then be observed by an optical telescope,
typically consisting of a collecting mirror and a camera. To properly
reconstruct the properties of such air showers, the atmospheric conditions at
the site have to be known in order to correct for Rayleigh scattering effects
and to estimate the fluorescence yield of the air
shower~\cite{Abraham:2010atmo}. Height-dependent profiles of temperature,
pressure and humidity as well as weather conditions near the ground are
relevant.

The \pao~\cite{Abraham:2004auger} is a cosmic ray detector located near \mal
in the Mendoza province in Argentina. It consists of a Surface Detector (SD)
array and five Fluorescence Detector (FD) buildings~\cite{Abraham:2009fd}.
Between 2002 and 2010, atmospheric conditions over the Observatory were measured
by intermittent meteorological radio soundings. Additionally, ground-based
weather stations measure surface data continuously in order to provide the
atmospheric parameters to properly reconstruct the measured air showers.

In south-east Colorado, several balloon soundings were performed as part of an
atmospheric \rnd project. The aim of this effort was to study possible
enhancements and performance improvements for the \pao, as well as explore
technological advancements for a possible future ground-based observatory. The
ground station used for the soundings was a mobile and slightly advanced version
of the equipment used in Argentina. The launches were performed at two sites,
the Atmospheric Monitoring Telescope (AMT) and the Distant Raman Laser Facility
(DRLF)~\cite{Wiencke:2011icrc}. The sites are about 40\,km apart and are both
equipped with identical weather stations.

Performing radio soundings imposes a large burden, both in terms of funds and
manpower. We investigated the possibility of using data from the Global Data
Assimilation System (GDAS)~\cite{GDASinformation}, a global atmospheric model,
for the site of the \pao~\cite{Abreu:2012gdas,Will:2011icrc}.  GDAS data are
publicly available free of charge via READY (Real-time Environmental
Applications and Display sYstem). Each data set contains all the main state
variables as a function of altitude. The data gathered in Colorado were also
compared to GDAS data in order to evaluate the possibility to use GDAS also in
different locations and for a possible future ground-based cosmic ray detector.

\section{Global Data Assimilation System\label{sec:gdas}}

\begin{figure*}[!t]
  \begin{minipage}[t]{.33\textwidth}
    \centering
    \includegraphics*[width=2.in]{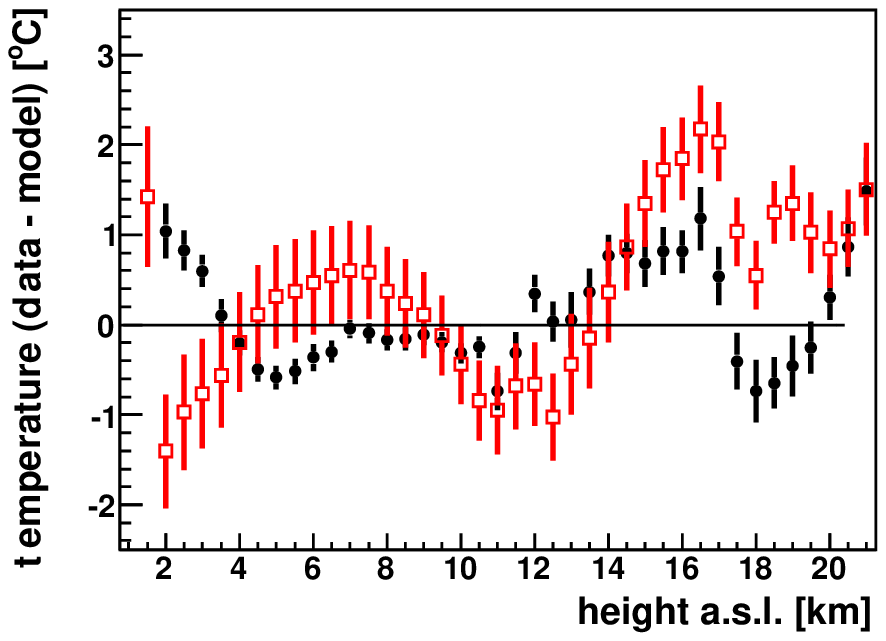}
  \end{minipage}
  \hfill
  \begin{minipage}[t]{.33\textwidth}
    \centering
    \includegraphics*[width=2.in]{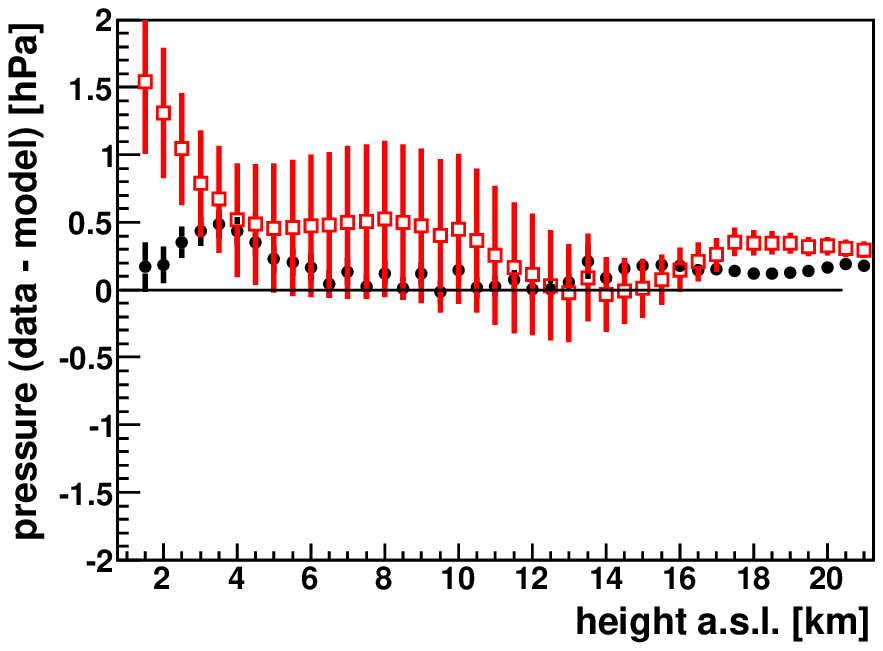}
  \end{minipage}
  \hfill
  \begin{minipage}[t]{.33\textwidth}
    \centering
    \includegraphics*[width=2.in]{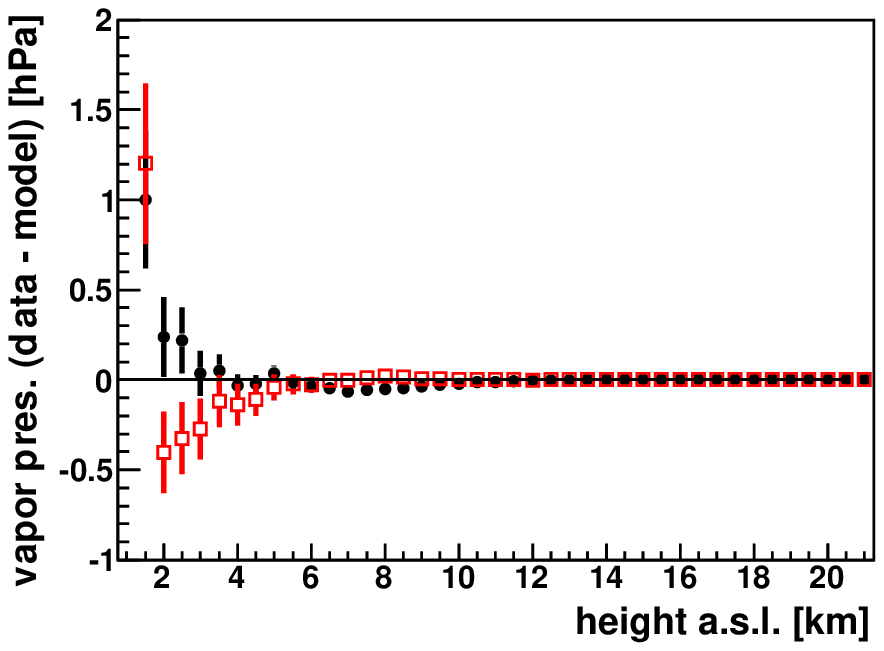}
  \end{minipage}
  \begin{minipage}[t]{.33\textwidth}
    \centering
    \includegraphics*[width=2.in]{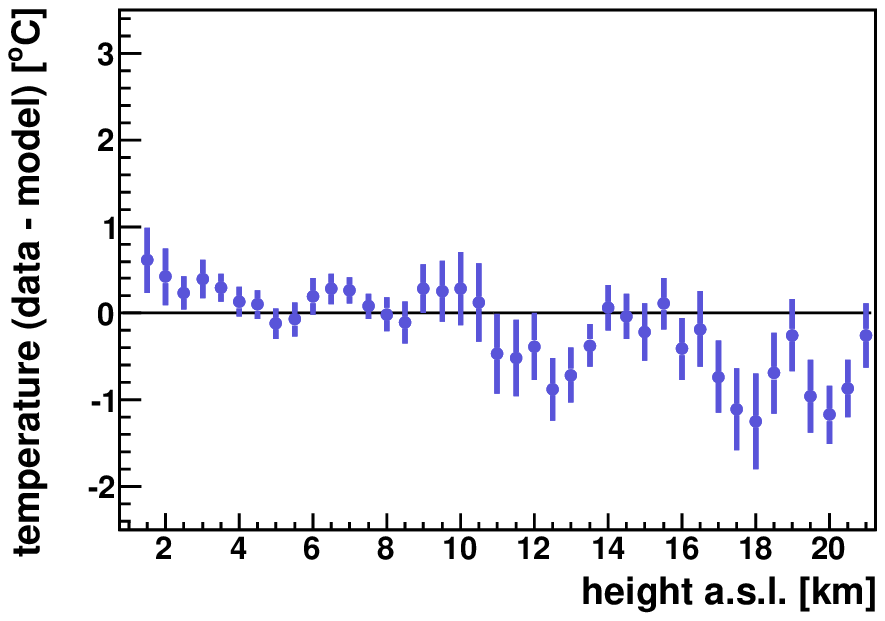}
  \end{minipage}
  \hfill
  \begin{minipage}[t]{.33\textwidth}
    \centering
    \includegraphics*[width=2.in]{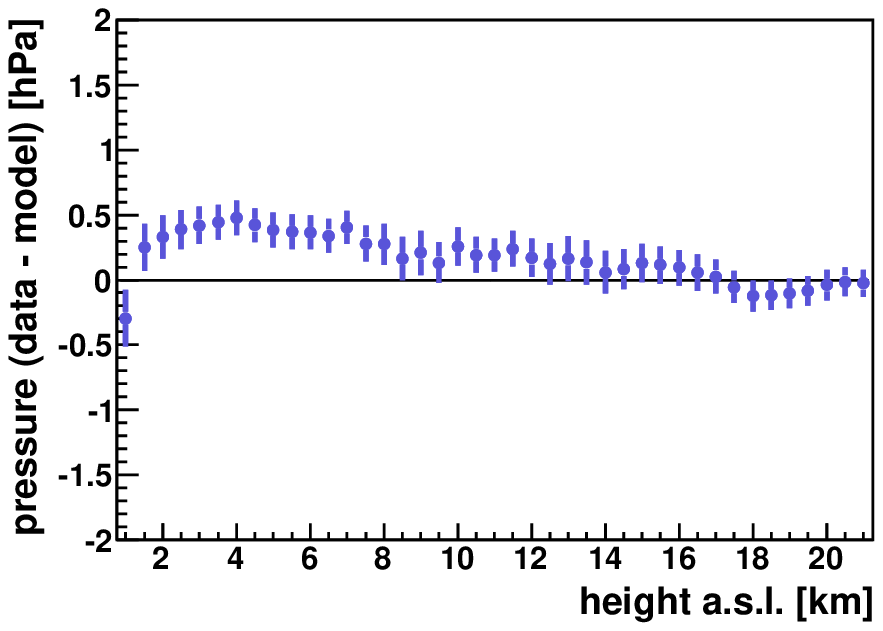}
  \end{minipage}
  \hfill
  \begin{minipage}[t]{.33\textwidth}
    \centering
    \includegraphics*[width=2.in]{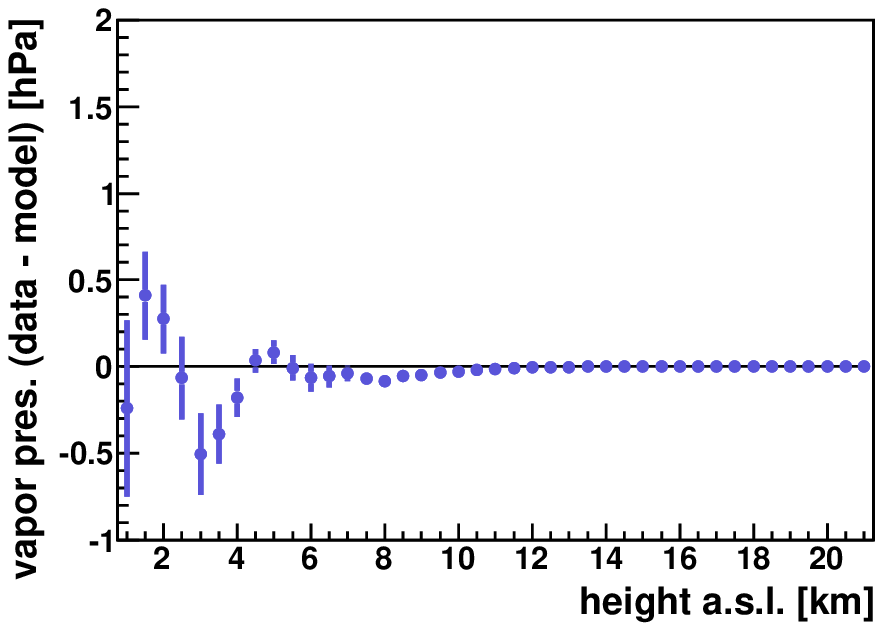}
  \end{minipage}
  \caption{\label{fig:SondeVsGDAS}
    \small{Top: Difference between measured individual radiosonde data and the
      corresponding GDAS data (black dots) and MM (red squares) versus height
      for all ascents performed at the \pao in Argentina in 2009 and 2010.
      Bottom: Difference between radiosonde data and the GDAS data (blue dots)
      versus height for all ascents performed at the Colorado \rnd site in 2009
      and 2010.}
  }
\end{figure*}

Data assimilation is a process in numerical weather prediction in which the
development of a model incorporates the real behavior of the atmosphere as found
in meteorological observations~\cite{Mueller:2004}. The atmospheric models
describe the atmospheric state at a given time and position. The first step in
performing a full data assimilation is to collect data from meteorological
instruments placed all over the world. Using the current atmospheric conditions,
a future state --~e.\,g. 3 hours ahead~-- is forecast using numerical weather
prediction. Finally, data assimilation is used to adjust the model output to the
measured atmospheric state, resulting in a 3-dimensional image of the
atmosphere. At a given time, the value of a state variable is known from
observations. For the same time, a model forecast for this variable from a
previous iteration a few hours earlier exists. The data assimilation step
combines observation and forecast. This analysis is the initial point for the
weather prediction model to create the forecast for a later time, when this
process is repeated.

The Global Data Assimilation System is an atmospheric model developed at the
National Centers for Environmental Prediction of the National Oceanic and
Atmospheric Administration. The numerical weather prediction model used is the
Global Forecast System. Data are available for every three hours at 23 constant
pressure levels --~from 1000\,hPa ($\approx$ sea level) to 20\,hPa ($\approx$
26\,km)~-- on a global 1\degree-spaced latitude-longitude grid (180\degree by
360\degree). Each data set is complemented by data for the surface. The data are
made available online~\cite{GDASinformation}.

For the site of the \pao, applicable GDAS data are available starting June 2005.
Because of the lateral homogeneity of the atmospheric variables across the Auger
array~\cite{Abraham:2010atmo}, one location is sufficient to describe the
atmospheric conditions. The grid point at 35\degree~S and 69\degree~W was
chosen, at the north-eastern edge of the Observatory. The grid point for the
Colorado \rnd site is 38\degree~N and 102\degree~W, about 40\,km to the east of
the DRLF and 60\,km to the north-east of the AMT. Since the terrain is very
similar to the Argentinian high desert, horizontal uniformity can be assumed.
This assumption was verified by radiosonde launches at different starting
positions.

For the air shower analyses of the \pao, the main state variables of the
atmosphere --~temperature, pressure and relative humidity~-- are needed at
several altitudes. They are provided directly by the GDAS surface and upper air
data. From those, air density and atmospheric depth profiles are calculated.

\section{GDAS vs.\ Measurements\label{sec:gdasVSlocal}}

To validate the quality of GDAS data and to verify their applicability to air
shower reconstructions for the \pao, we compare the GDAS data with local
soundings from weather balloons and ground-based weather stations. Comparisons
using the data from the Colorado site are also shown.

\subsection{GDAS vs.\ Weather Balloon Soundings\label{sec:GDASvsRadio}}

Local radio soundings are performed above the array of the \pao since 2002, but
not on a regular basis. To provide a set of atmospheric data for every measured
event, the profiles from the ascents were averaged to obtain local models,
called \mal Monthly Models (MM)~\cite{Will:2012epj}. The MM have been
compiled using data until the end of 2008. The uncertainties for each variable
are given by the standard deviation of the differences within each month
together with the absolute uncertainties of the sensors measuring the
corresponding quantity.

Comparing the monthly models with ascent data until the end of 2008 shows, by
construction, only small deviations~\cite{Abraham:2010atmo}. In the comparison
displayed in the top panels of Fig.~\ref{fig:SondeVsGDAS}, radiosonde data from
2009 and 2010 are used to illustrate the strength of the GDAS data, the
data set of local soundings being independent of the MM. The error bars denote
the RMS of the differences at each height. These uncertainties are larger for
the MM than for GDAS data. In contrast, the GDAS data represent the local
conditions in 2009 and 2010 much better and the intrinsic uncertainty is
consistently small. For earlier years, the GDAS data fit the measured data
equally well or better than the MM which were developed using the data from
these years.

In the bottom panels of Fig.~\ref{fig:SondeVsGDAS}, the same comparison is shown
between the radiosonde data measured in Colorado and the corresponding GDAS
data. The differences are of the same order and the error bars are similar to
the results at the \pao site.

The GDAS data fit the radiosonde data in the upper part of the atmosphere,
especially in the field of view of the fluorescence detectors. Possible
inconsistencies between local measurements and GDAS data close to the ground are
investigated using weather station data.

\subsection{GDAS vs.\ Ground Weather Stations\label{sec:GDASvsWS}}

Five ground weather stations continuously monitor atmospheric values at the
\pao. They are mounted between about 2 to 5\,m above ground level at four FD
stations, and one was set up near the center of the array at the Central Laser
Facility (CLF). For the Colorado \rnd site, two identical weather stations were
set up, one at the DRLF laser facility and one at the AMT telescope site. To
make sure that the GDAS data describe the conditions at the ground reasonably
well, the values provided by the GDAS data set are compared to all available
weather station data. The profiles built using GDAS data are interpolated at the
height of the station.

\begin{figure}[!t]
  \includegraphics*[width=.49\linewidth]{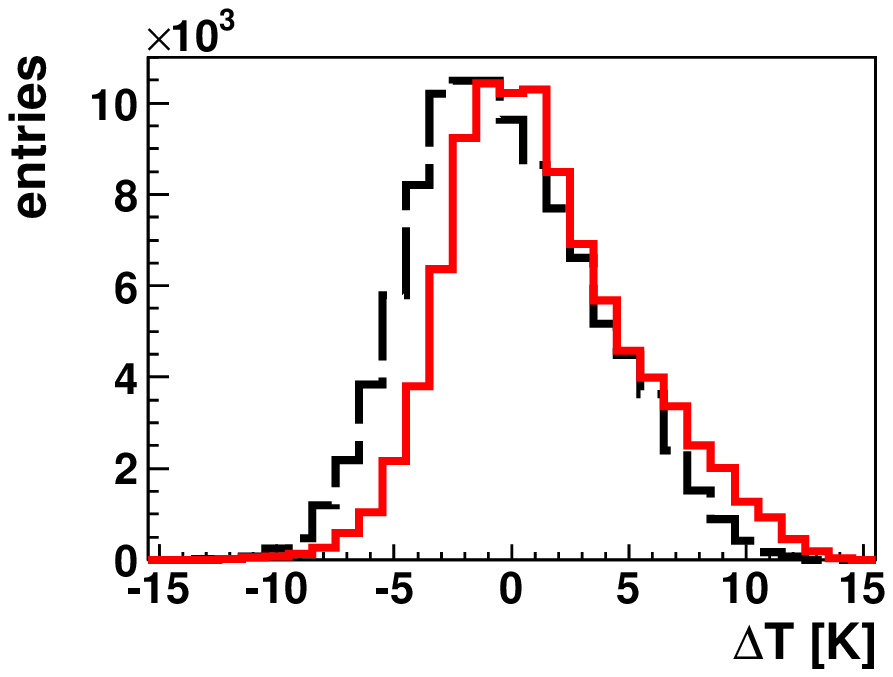}
  \hfill
  \includegraphics*[width=.49\linewidth]{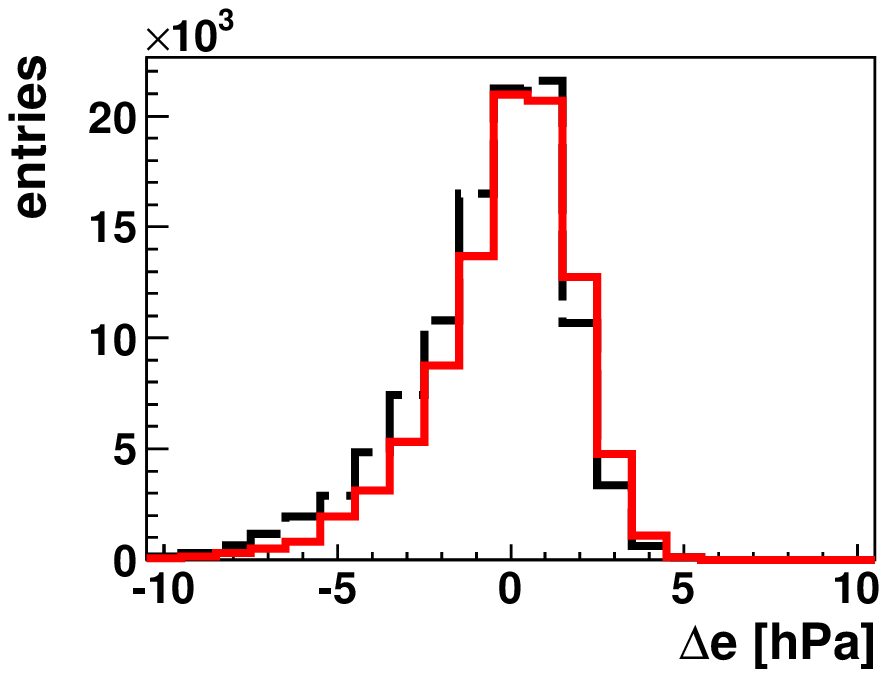}
  \includegraphics*[width=.49\linewidth]{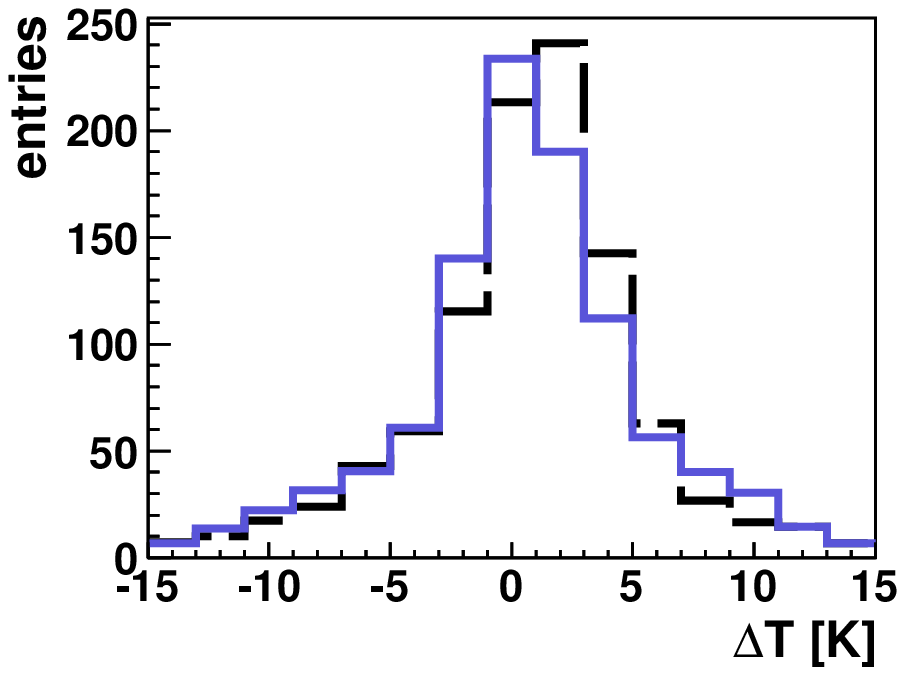}
  \hfill
  \includegraphics*[width=.49\linewidth]{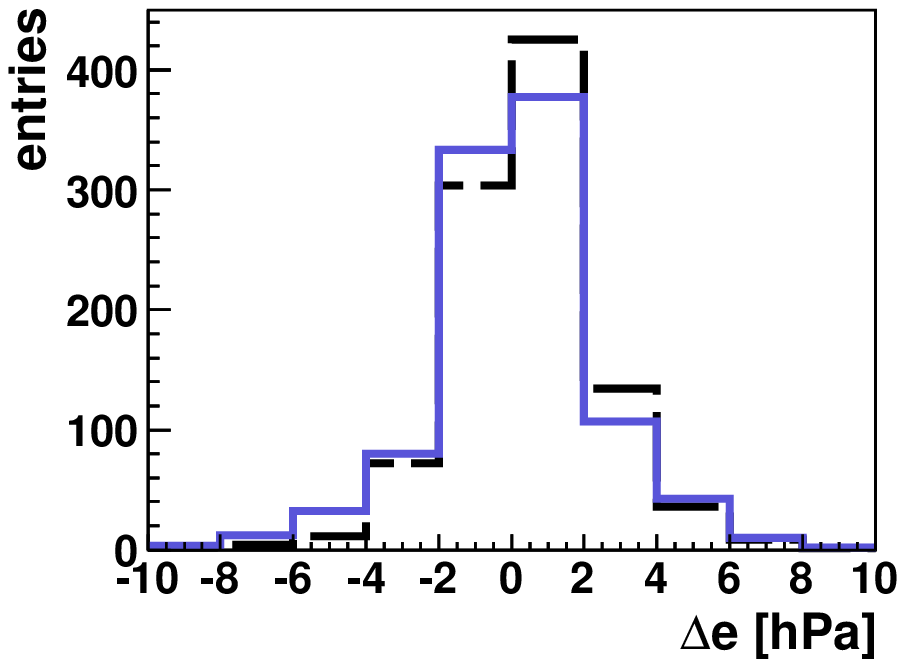}
  \caption{\label{fig:WSvsGDAS}
    \small{Difference between data measured at weather stations and from GDAS
    (`GDAS' minus `weather station') in temperature and water
    vapor pressure are shown. Top: Data from 2009 of the CLF (dashed line) and
    Loma Amarilla (solid line) stations at the \pao. Bottom: Data taken between
    January 2010 and June 2011 at the AMT (dashed line) and DRLF (solid line)
    stations in Colorado.}
  }
  \hfill
\end{figure}

In Fig.~\ref{fig:WSvsGDAS}, the differences between measured weather station
data and GDAS data are shown. For the stations at the CLF and the FD site Loma
Amarilla (LA) all data measured in 2009 were used (top panel). For the Colorado
\rnd site, data taken between January 2010 and June 2011 were used for both the
AMT and DRLF sites. Temperature, pressure (not shown), and vapor pressure are in
similar agreement as GDAS data with local sounding data close to ground (cf.\
Fig.~\ref{fig:SondeVsGDAS}). The mean difference in temperature is 1.3\,K for
the CLF, $-$0.3\,K for the LA, 0.5\,K for the DRLF and 0.7 for the AMT station.
For vapor pressure, the means are $-$0.2\,hPa (CLF), $-$0.7\,hPa (LA), 0.2\,hPa
(DRLF) and 0.4\,hPa (AMT). The differences between the GDAS and the weather
station data are of the same order as the difference in data of two different
stations~\cite{Abreu:2012gdas}.

The GDAS data fit the measured data at the Observatory and the \rnd site very
well and are better suited for use in air shower reconstructions and simulation
than monthly mean models. This reduces the need for laborious and costly
radiosonde launches to sporadic checks of the consistency of the GDAS data.

\section{Air Shower Reconstruction\label{sec:reco}}

To study the effects caused by using GDAS data in the air shower reconstruction
of the \pao, all air shower data between June~1, 2005 and the end of 2010 were
used. The change of the description of the atmosphere will mainly affect the
reconstruction of the fluorescence data. Varying atmospheric conditions alter
the fluorescence light production and transmission~\cite{Abraham:2010atmo}. The
fluorescence model we use determines the fluorescence light as a function of
atmospheric conditions~\cite{Keilhauer:2010}, parameterized using results from
the AIRFLY experiment~\cite{Ave:2007,Ave:2008}.

\subsection{Data Reconstruction}

\begin{figure}[!t]
  \includegraphics*[width=.49\linewidth]{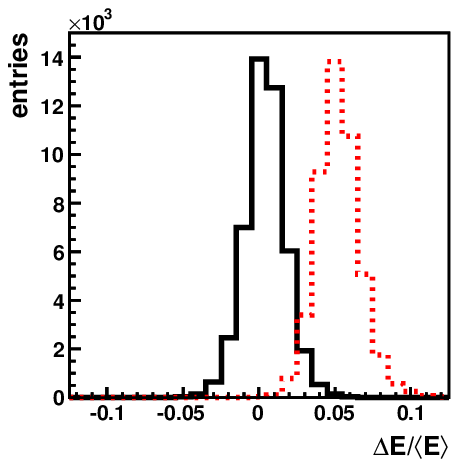}
  \hfill
  \includegraphics*[width=.49\linewidth]{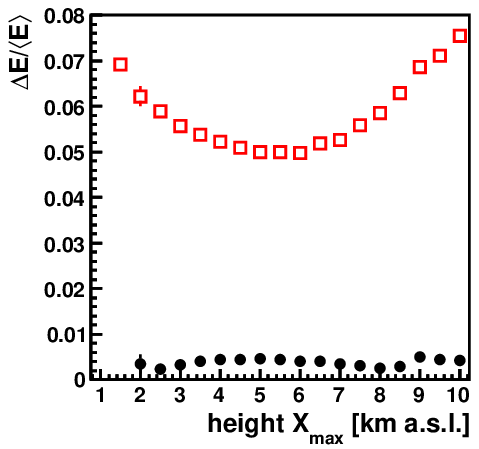}
  \includegraphics*[width=.49\linewidth]{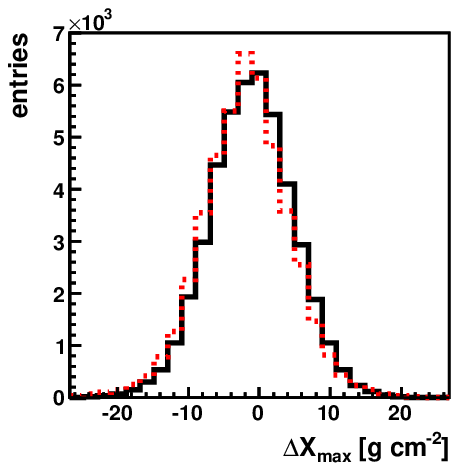}
  \hfill
  \includegraphics*[width=.49\linewidth]{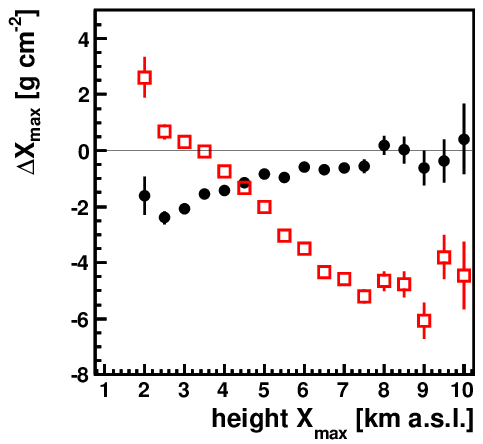}
  \caption{\label{fig:delta}
    \small{Difference of reconstructed $E$ (top) and \xmax (bottom), plotted
    versus geometrical height of \xmax in the right panels. Dashed black line or
    black dots for \textsf{FY$_{\rm mod}^{\rm GDAS}$} minus \textsf{FY$_{\rm
    mod}$}, and solid red line or open red squares for \textsf{FY$_{\rm
    mod}^{\rm GDAS}$} minus \textsf{FY}.}
  }
\end{figure}

The following analysis is based on three sets of reconstructions. The first set,
\textsf{FY}, is the reconstruction applying an atmosphere-dependent fluorescence
yield calculation without temperature-dependent collisional cross sections and
humidity quenching~\cite{Arqueros:2008}. The MM are used in the calculations.
For the second set, \textsf{FY$_{\rm mod}$}, all atmospheric effects in the
fluorescence calculation are taken into account. Again, the MM are used. For the
third set, \textsf{FY$_{\rm mod}^{\rm GDAS}$}, the MM are exchanged with the new
GDAS data in combination with the modified fluorescence calculation. Comparing
the reconstruction sets with each other, the variation of the reconstructed
primary energy $E$ and the position of shower maximum \xmax can be determined,
see Fig.~\ref{fig:delta}.

Using GDAS data in the reconstruction instead of MM affects $E$ only slightly.
The mean of the difference \textsf{FY$_{\rm mod}^{\rm GDAS}$} minus
\textsf{FY$_{\rm mod}$} is 0.4\% with an RMS of 1.4\%. For the reconstructed
\xmax, only a small shift of $-$1.1\,\gcm is found with an RMS of 6.0\,\gcm.
Comparing the full atmosphere-dependent reconstruction \textsf{FY$_{\rm
mod}^{\rm GDAS}$} with \textsf{FY}, a clear shift in $E$ can be seen: an
increase in $E$ by 5.2\% (RMS 1.5\%) and a decrease of \xmax by $-$1.9\,\gcm
(RMS 6.3\,\gcm). These modified fluorescence settings are now used in the Auger
reconstruction, in conjunction with other improvements to the procedure,
see~\cite{Verzi:2013}.

The description of atmospheric conditions close to ground is very difficult in
monthly mean profiles since the fluctuations in temperature and humidity are
larger below 4\,km than in the upper layers of the atmosphere. Consequently, a
more precise description of actual atmospheric conditions with GDAS than with MM
will alter the reconstruction for those air showers which penetrate deeply into
the atmosphere. The full atmosphere-dependent fluorescence calculation alters
the light yield for conditions with very low temperatures, corresponding to
higher altitudes. Showers reaching their maximum in the altitude range between 3
and 7\,km show a difference in $E$ around 5\%, see Fig.~\ref{fig:delta}, upper
right. However, showers with very shallow or very deep \xmax are reconstructed
with a 7--8\% higher energy than using the atmosphere-independent fluorescence
calculation. The \xmax sensitivity to the different parameterizations of the
atmosphere and fluorescence yield (Fig.~\ref{fig:delta}, lower right) is
consistent to what has been reported in~\cite{Keilhauer:2008}.

\subsection{Impact on Reconstruction Uncertainties}

To study the effect that GDAS data have on the uncertainties of air shower
reconstructions, air showers induced by protons and iron nuclei are simulated
with energies between 10$^{17.5}$\,eV and 10$^{20}$\,eV. The fluorescence light
is generated using temperature-dependent cross sections and water vapor
quenching. The times of the simulated events correspond to 109~radio soundings
between August 2002 and December 2008 so that realistic atmospheric profiles can
be used in the simulation. All launches were performed at night during
cloud-free conditions. After the atmospheric transmission, the detector optics
and electronics are simulated. The resulting data are reconstructed using the
radiosonde data, as well as the GDAS data.

Some basic quality cuts are applied to the simulated showers. The same study has
been performed to determine the uncertainties of the MM~\cite{Keilhauer:2009}.
The systematic error due to different atmospheres was found to be less than 1\%
in $E$ and less than 2\,\gcm in \xmax. Between 10$^{17.5}$\,eV and
10$^{20}$\,eV, energy-dependent reconstruction uncertainties of $\pm$1\% and
$\pm$5\,\gcm for low energies and up to $\pm$2\% and $\pm$7\,\gcm for high
energies were found.

\begin{figure}[!t]
  \includegraphics*[width=.49\linewidth]{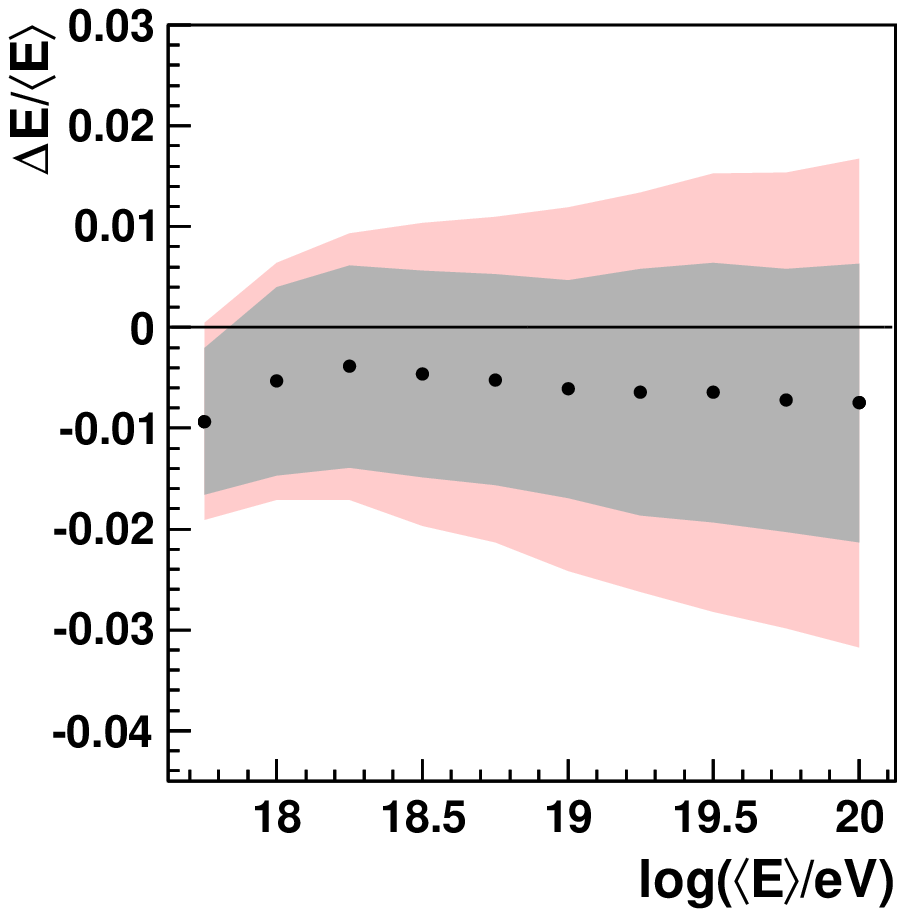}
  \includegraphics*[width=.49\linewidth]{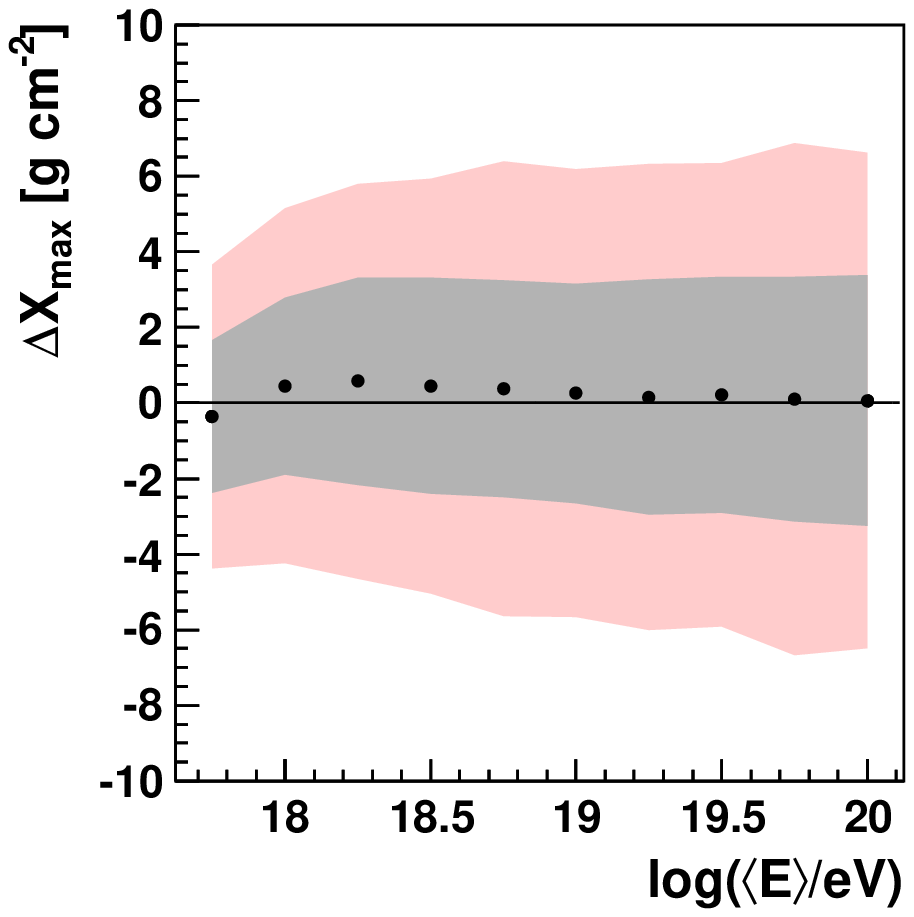}
  \caption{\label{fig:sim_uncert}
    \small{Energy difference (left) and \xmax difference (right) vs.\
    reconstructed FD energy for simulated showers. Gray bands denote the true
    RMS spread for the GDAS reconstructions, the red band indicates the RMS for
    the reconstructions using monthly models.}
  }
  \hfill
\end{figure}

In Fig.~\ref{fig:sim_uncert}, the influence on the reconstruction due to GDAS
data is shown. A deviation from zero indicates a systematic error, the gray
error bands denote the true RMS spread of all simulated events and are a measure
of the reconstruction uncertainty due to the atmospheric parameterization using
GDAS. The red bands indicate the same RMS spread for the reconstructions using
the MM. The systematic shifts in $E$ are below 1\%, and the shifts in \xmax are
less than 0.5\,\gcm. The RMS spread for GDAS is considerably smaller than for
the MM, $\pm$0.9\% and $\pm$2.0\,\gcm for low energies, $\pm$1.3\% and
$\pm$3.5\,\gcm for high energies. The $E$ uncertainty at low energies is
comparable to that introduced by the MM. At high energies, the uncertainty is
almost half. For \xmax, the uncertainties at all energies are halved.

This study of the reconstruction uncertainties using different atmospheric
parameterizations further demonstrates the advantages of GDAS data over the MM.

\section{Conclusion\label{sec:conclusion}}

The comparison of GDAS data for the site of the \pao in Argentina with local
atmospheric measurements validated the adequate accuracy of the 3-hourly GDAS
data. An air shower reconstruction analysis confirmed the applicability of GDAS
for Auger reconstructions and simulations, giving improved accuracy when
incorporating GDAS data instead of MM. Also, the value of using an
atmosphere-dependent fluorescence description has been demonstrated. For the
Colorado \rnd site, the differences between the measured radiosonde data and
GDAS are of the same order as in Argentina, further supporting the general
validity of GDAS data as an atmospheric description to be used in current and
future cosmic ray observatories.

\vspace*{0.5cm}
\footnotesize{{\bf Acknowledgment:}
We would like to thank the organizers of the workshop \emph{AtmoHEAD:
Atmospheric Monitoring for High-Energy Astroparticle Detectors} in Saclay,
France, 2013 for the inspiring meeting. Part of these investigations are
supported by the Bundesministerium f\"ur Bildung und Forschung (BMBF) under
contracts 05A08VK1 and 05A11VK1. Furthermore, these studies would not have been
possible without the entire Pierre Auger Collaboration and the local staff of
the \pao.{}}

\clearpage

\end{document}